\documentclass[a4paper,12pt, epsfig]{article}
\usepackage{epsfig}
\usepackage{amssymb}
\usepackage{amsfonts}

\newskip\humongous \humongous=0pt plus 1000pt minus 1000pt

\newif\ifdtup

\catcode`\@=11

\@addtoreset{equation}{section}
\def\theequation{\thesection.\arabic{equation}}

\def\@normalsize{\@setsize\normalsize{15pt}\xiipt\@xiipt
\abovedisplayskip 14pt plus3pt minus3pt%
\belowdisplayskip \abovedisplayskip
\abovedisplayshortskip \z@ plus3pt%
\belowdisplayshortskip 7pt plus3.5pt minus0pt}

\def\small{\@setsize\small{13.6pt}\xipt\@xipt
\abovedisplayskip 13pt plus3pt minus3pt%
\belowdisplayskip \abovedisplayskip
\abovedisplayshortskip \z@ plus3pt%
\belowdisplayshortskip 7pt plus3.5pt minus0pt
\def\@listi{\parsep 4.5pt plus 2pt minus 1pt
      \itemsep \parsep
      \topsep 9pt plus 3pt minus 3pt}}

\relax

\catcode`@=12

\catcode`\@=11

\def\section{\@startsection{section}{1}{\z@}{3.5ex plus 1ex minus
    .2ex}{2.3ex plus .2ex}{\large\bf}}

\def\thesection{\arabic{section}}
\def\thesubsection{\arabic{section}.\arabic{subsection}}

\def\appendix{\setcounter{section}{0}
  \def\thesection{Appendix \Alph{section}}
  \def\thesubsection{\Alph{section}.\arabic{subsection}}
  \def\theequation{\Alph{section}.\arabic{equation}}}

\def\SymBoxes#1#2#3#4{\newdimen\un@t \un@t#3%
\raisebox{#1}{\rule{#2\un@t}{#4}\hskip-#2\un@t
\@tempdimb\un@t \advance\@tempdimb by-#4\@tempcntb#2\relax%
\@whilenum{\@tempcntb>0}\do{
\rule{#4}{\un@t}\hskip\@tempdimb \advance\@tempcntb by\m@ne}%
\hskip-#2\un@t \rule[\un@t]{#2\un@t}{#4}%
\rule[\un@t]{#4}{#4}\hskip-#4
\rule{#4}{\un@t}}\hskip-#4}                

\begin{document}

\newcommand{\beq}{\begin{equation}}
\newcommand{\eeq}{\end{equation}}
\newcommand{\bea}{\begin{eqnarray}}
\newcommand{\eea}{\end{eqnarray}}
\newcommand{\beas}{\begin{eqnarray*}}
\newcommand{\eeas}{\end{eqnarray*}}
\newcommand{\defi}{\stackrel{\rm def}{=}}
\newcommand{\non}{\nonumber}
\newcommand{\bquo}{\begin{quote}}
\newcommand{\enqu}{\end{quote}}
\renewcommand{\(}{\begin{equation}}
\renewcommand{\)}{\end{equation}}
\def\de{\partial}
\def\Om{\ensuremath{\Omega}}
\def\Tr{ \hbox{\rm Tr}}
\def\H{ \hbox{\rm H}}
\def\HE{ \hbox{$\rm H^{even}$}}
\def\HO{ \hbox{$\rm H^{odd}$}}
\def\HEO{ \hbox{$\rm H^{even/odd}$}}
\def\HOE{ \hbox{$\rm H^{odd/even}$}}
\def\HHO{ \hbox{$\rm H_H^{odd}$}}
\def\HHEO{ \hbox{$\rm H_H^{even/odd}$}}
\def\HHOE{ \hbox{$\rm H_H^{odd/even}$}}
\def\K{ \hbox{\rm K}}
\def\Im{ \hbox{\rm Im}}
\def\Ker{ \hbox{\rm Ker}}
\def\const{\hbox {\rm const.}}
\def\o{\over}
\def\im{\hbox{\rm Im}}
\def\re{\hbox{\rm Re}}
\def\bra{\langle}\def\ket{\rangle}
\def\Arg{\hbox {\rm Arg}}
\def\Re{\hbox {\rm Re}}
\def\Im{\hbox {\rm Im}}
\def\exo{\hbox {\rm exp}}
\def\diag{\hbox{\rm diag}}
\def\longvert{{\rule[-2mm]{0.1mm}{7mm}}\,}
\def\a{\alpha}
\def\dag{{}^{\dagger}}
\def\tq{{\widetilde q}}
\def\p{{}^{\prime}}
\def\W{W}
\def\N{{\cal N}}
\def\hsp{,\hspace{.7cm}}
\newcommand{\C}{\ensuremath{\mathbb C}}
\newcommand{\Z}{\ensuremath{\mathbb Z}}
\newcommand{\R}{\ensuremath{\mathbb R}}
\newcommand{\rp}{\ensuremath{\mathbb {RP}}}
\newcommand{\cp}{\ensuremath{\mathbb {CP}}}
\newcommand{\vac}{\ensuremath{|0\rangle}}
\newcommand{\vact}{\ensuremath{|00\rangle}                    }
\newcommand{\oc}{\ensuremath{\overline{c}}}
\begin{titlepage}
\begin{flushright}
SISSA 19/2008/EP
\end{flushright}
\bigskip
\def\thefootnote{\fnsymbol{footnote}}

\begin{center}
{\large {\bf
Topological strings live on attractive manifolds
  } }
\end{center}

\bigskip
\begin{center}
{\large  Jarah Evslin$^{1}$\footnote{\texttt{evslin@sissa.it}} and Ruben
Minasian$^{2}$\footnote{\texttt{ruben.minasian@cea.fr}}}
\end{center}

\renewcommand{\thefootnote}{\arabic{footnote}}

\begin{center}
\vspace{1em}
{\em  $^1${ SISSA,\\
Via Beirut 2-4,\\
I-34014, Trieste, Italy\\
\vskip .4cm
$^2$ Institut  de Physique Th\'eorique,
CEA/Saclay \\
91191 Gif-sur-Yvette Cedex, France  }
}
\end{center}

\noindent
\begin{center} {\bf Abstract} \end{center}
We add to the mounting evidence that the topological B model's normalized holomorphic three-form has integral periods by demonstrating that otherwise the B2-brane partition function is ill-defined.  The resulting Calabi-Yau manifolds are roughly fixed points of attractor flows.  We propose here that any admissible background for topological strings requires a quantized (twisted) integrable pure spinor, yielding a quantized (twisted) generalized Calabi-Yau structure. This proposal would imply in particular that the A model is consistent only on those Calabi-Yau manifolds that correspond to melting crystals.  When a pure spinor is not quantized, type change occurs on positive codimension submanifolds. We find that quantized pure spinors in topological A-model instead change type only when crossing a  coisotropic 5-brane.
Quantized generalized Calabi-Yau structures do correspond to twisted K-theory classes, but some twisted K-theory classes correspond to either zero or to multiple structures.


\vfill

\begin{flushleft}
{\today}
\end{flushleft}
\end{titlepage}

\hfill{}


\setcounter{footnote}{0}
\section{Introduction}
It is hoped that topological string theories hold the key to understanding quantum gravity.  For the time being they provide a powerful calculational simplification of string theory. When considered on internal spaces that preserve a sufficient amount of supersymmetry they compute quantities which agree with those of physical string theories compactified on the same space.     For example, given a manifold $M$ one may use topological string theory to calculate not only its geometric properties, like Gromov-Witten invariants, but also to calculate physical quantities such as the prepotential of an $\N$=2 gauge theory obtained by compactification on $M$ or the superpotential of an $\N$=1 compactification.  As the A model partition function is roughly equal to the elliptic genus at $q=e^{-g_s}$, which counts BPS microstates of 5d black holes with fixed spin, one can even use topological string theory to calculate black hole entropies \cite{9809187,9812127,0410178}.  Is every $M$ yielding a supersymmetric string background admissible as a topological string theory background?

A necessary condition for supersymmetric compactifications of physical strings is that $M$ is a (twisted) generalized Calabi-Yau space, i.e. it admits a globally-defined (twisted) closed pure spinor \cite{gmpt1, gmpt2}. These are special polyforms whose lowest degree part is a good characterization of the geometry \cite{math/0209099, math/0401221}.  We will refer to the lowest degree as the type of the generalized complex structure.  Preserving supersymmetry in Type II compactifications puts further constraints on $M$. A compatible pure spinor must exist. When the latter is also closed, we are dealing with ordinary Calabi-Yau manifolds; when not - this failure of integrability fixes the RR sector modulo the tadpole constraints.

Calabi-Yau manifolds are generally considered to be admissible topological backgrounds. Moreover at least at a formal level, on manifolds with a trivial canonical bundle a definition of an A or a  B model can be given using only the symplectic or the complex structure respectively. These are nicely described by closed pure spinors (of type 0 or 3 respectively) . New more general proposals have appeared suggesting that a single  generic (twisted) pure spinor, in other words a generalized Calabi-Yau structure, suffices to define a topological model.  We will not attempt to assess whether such models can be defined, but rather we assume that they exist and we try to understand the consequences of the constraints placed by the well-definedness of the topological brane partition functions.

In this note we restrict this proposal somewhat, by suggesting  that topological string theory backgrounds must be (twisted) generalized Calabi-Yau's whose pure spinors are necessarily (twisted) quantized.  In particular, this eliminates all but a discrete, although sometimes dense, subset of the moduli space of ordinary Calabi-Yau's and so all but a measure zero set of known backgrounds.

\subsection{Indirect evidence for quantization}

Circumstantial evidence for the quantization of the holomorphic three-form $\Om$ in the B model topological string theory has been accumulating for years.  In section 5 of Ref.~\cite{0305132} the authors propose that the integral of $\Om$ over a contractible 3-cycle is quantized and is equal to
\beq
\int\Om=g_sN
\eeq
where $g_s$ is the string coupling and $N$ is the linking number with B2-branes, which wrap holomorphic curves in the B model on a Calabi-Yau.  This condition is equivalent to the Bianchi identity
\beq
d\Om=g_s\delta^4(Q) \label{bianchi1}
\eeq
where $Q$ is the weighted union of the worldvolumes of the B2-branes.  As interpreted in Ref.~\cite{0312022}, a similar condition was found in section 2 of Ref.~\cite{0305132} for the K\"ahler form $K$ on a contractible 2-cycle in the A model
\beq
\int K=g_sN
\eeq
where now $N$ is the number of Lagrangian A3-branes linked by the two cycle.

In the case of physical string theories, when a quantization condition is found for fluxes on a contractible cycle, one can arguably extend it to fluxes on noncontractible cycles by stating that maybe it is contractible far away and the physics cannot depend on such nonlocal information, as in Ref.~\cite{9609122}.  In a topological string theory one cannot deform the space as one likes and arguments based on the metric are suspect.

Further evidence for the quantization of the K\"ahler form $K$ on noncontractible cycles in the A model and $\Om$ on noncontractible cycles in the B model came from their relations to other problems.  In Ref.~\cite{0309208} the authors found that the A model is equivalent to a three-dimensional melting crystal.  This was interpreted in Ref.~\cite{0312022} as a quantum foam description of A model physics.  Interestingly, the K\"ahler potential divided by $g_s$ is a Chern class in this description, and so the equivalence only applies when the K\"ahler potential is quantized.  This on its own does not demonstrate that the K\"ahler potential is always quantized, but merely that it is quantized in the cases in which the A model is equivalent to a melting crystal.

In the B model when $\Om$ is quantized, the backgrounds are attractive manifolds \cite{9807087}, which are the near horizon geometries of fixed points of attractor flows in the physical string theory.  While it is not clear how to interpret split flows in this framework, this connection provides a geometric interpretation for Calabi-Yau's with integer periods, even allowing a natural extension to the generalized complex case where the discriminant is the square of Hitchin's functional.  Using the conjectures of Ref.~\cite{9807087} one may even find a correspondence between allowed B model backgrounds and rational CFTs, conclude that background varieties must admit complex multiplication, or even hope that backgrounds which satisfy the quantization condition are dense, as, at least in some neighborhood, is the case for the compactifications on $T^6$ and $K3\times T^2$ considered in that note.  This would be in contrast with the quantization of the Maxwell field or fluxes in physical string theory, which give an integral lattice which is by no means dense.  The density of the attractive manifolds heavily relies on the automorphisms of the spaces considered, which map the lattice into a compact fundamental domain and thus allow it to be dense, but it is not known if such large automorphism groups are generic among attractive Calabi-Yau's.  Needless to say, while the connection to attractors is intriguing it is by no means a demonstration that $\Om$ is quantized.

\subsection{Quantization from holomorphic Chern-Simons}

We will now present what we consider to be much less circumstantial evidence than the arguments reviewed above for the quantization of the periods of $\Om$.  Before any of the above evidence was collected, Aganagic and Vafa dimensionally reduced the holomorphic Chern-Simons theory on the spacefilling B6 to find a singularly simple formula for the action of the B2-brane \cite{0012041}.  In Section 4 of their paper they conclude that the B2-brane action is the inverse of the exterior derivative of the holomorphic three-form
\beq
S=\int_{C^2} g_s\alpha\hsp \Om=d\alpha
\eeq
where $C^2$ is the holomorphic curve wrapped by the B2.  They claim that while $\alpha$ is ill-defined, this integral is well-defined when $C^2$ has no boundary.

Perhaps there is a canonical way to render it well-defined, but the standard method of defining such an integral would be to define a 3-chain $D^3$ such that $C^2=\partial D^3$ and then to use Stokes' theorem to identify the action with the integral of $\Om$ on $D^3$.  One then needs to check that the action is independent of the choice of $D^3$.  Two different choices of $D^3$ can differ by a representative $Z^3$ of any element of $H_3(M,\Z)$ and so the difference in the actions is the integral of $\Om$ over $Z^3$.  One does not really need the difference in the actions to be zero, but really only the difference in the exponentials must vanish, and so one arrives at
\beq
1=\frac{e^{iS_1}}{e^{iS_2}}=\exp^{ig_s\int_{Z^3}\Om}
\eeq
which implies that the periods of $\Om$ are integral.  It may be, if one combines the B and $\overline{\rm B}$ models as suggested in Ref.~\cite{0411073}, that in keeping with the attractor analogy only the real part of $\Om$ appears in the action and so is quantized.  We will adopt this point of view.  While in the physical $\N$=2 flux compactifications the overall phase of the periods of $\Om$ can be rotated freely via an R-symmetry transformation, in the topological string theory it appears to be part of the data of the theory.

In practice it often happens that the quantization of the periods of a form on all of the cycles simultaneously is too much to ask, and it suffices to choose a polarization, choosing half of the cycles on which to quantize and then showing that the observables do not depend on which half is chosen.  For example, one may choose a Darboux basis for the 3-cycles of a Calabi-Yau and then quantize the periods on the electric cycles.  There is already a choice of polarization in the B-model implicit in the original choice of basepoint of the complex structure \cite{9306122}, which perhaps provides the natural polarization to use here.

In this paper we will formally extend the quantization of $\Om$ to all generalized complex structures, and discuss its possible implications for topological string theories.  In particular, in Ref.~\cite{0312022} a very similar argument to that above has been used to claim that the A1-brane partition function implies the quantization of the K\"ahler form.  While we have no arguments that suggest either the existence of an A1-brane in the A model or that its action has the form claimed by the authors, and in particular we know of no candidate 1-dimensional complex submanifolds on which it may be wrapped, the resulting quantization of the K\"ahler form will be the hypothesis from which the rest of our considerations of the A model follow.

In section \ref{hitchsec} we will discuss matter couplings to the Hitchin functional and argue in section \ref{ksec} that all allowed backgrounds correspond to twisted K-theory classes.  This leads naturally to a kind of Chern-Simons\footnote{We thank Giulio Bonelli for discussions on this point.} action on coisotropic A5-branes.  In section \ref{jumpsec} we describe type change in the A and B models, finding surprisingly that in the A model type 0 and type 2 regions must be separated by domain walls, which are coisotropic branes, unlike the predictions of Gualtieri's general analysis in the nonquantized case \cite{math/0401221} which would suggest that, as in the B model, higher types live on submanifolds inside regions of lower type.

\section{Hitchin's functional with branes} \label{hitchsec}

\subsection{The action and equations of motion}

According to the proposal of Ref.~\cite{GS,0411073,Nekrasov}, the topological data of the A model on a generalized Calabi-Yau $M$ is given by an even polyform $\phi$ while that of the B model is given by an odd polyform $\phi$.  More precisely, the theory that they consider has a partition function which appears to be the norm squared of that of the usual topological theory.  In fact the holomorphic anomaly suggests that perhaps the B model alone does not constitute a consistent theory.  Proposed formulations of topological strings on the corresponding generalized complex manifolds have appeared in Refs.~\cite{gmpt2,0407249, Zucchini, 0603145}.   The action is equal to the Hitchin functional \cite{math/0010054}
\beq
S=\int_M \phi\wedge\overline{\phi} \label{hitch}
\eeq
where $\overline{\phi}$ is the complex conjugate of $\phi$ with an extra minus sign in every fourth dimension, corresponding to every other term.  We have suppressed factors of $g_s$, which can be used to scale $\Om$ and the K\"ahler form $K$.  The integrability of the generalized Calabi-Yau structure implies
\beq
d\phi=0
\eeq
and so locally one may define a potential $\alpha$ by
\beq
\phi=d\alpha.
\eeq

As is usual in $p$-form gauge theories, electrically charged matter $j$ is coupled to the potential $\alpha$
\beq
S=\int_M \phi\wedge\overline{\phi}+\alpha\wedge j
\eeq
and so the equation of motion obtained by varying $\alpha$ is
\beq
d\overline{\phi}=j. \label{bianchi}
\eeq
We will identify the Poincar\'e dual of the current $j$ with the topological brane worldvolume.  {\it{Thus topological branes are the obstructions to the integrability of the generalized Calabi-Yau structure.}}  In this sense, they play a role similar to that of the RR field strengths in the physical string theories, which are obstructions to the integrability of one of the pure spinors in the generalized K\"ahler structure \cite{gmpt1,gmpt2}.  A crucial difference between the two obstructions is that the RR fluxes are smeared out and so in general the nonintegrable pure spinor in the physical theory is almost nowhere closed and so cannot be quantized.  Luckily we have no evidence to suggest the quantization of the integrable pure spinor in the physical theory.

In the case of the B model on a Calabi-Yau $\phi=\Om$ and so Eq.~(\ref{bianchi}) reproduces Eq.~(\ref{bianchi1}), where $j$ is a delta function sourced by a B2-brane worldvolume.  In the case of the A model on a symplectic manifold $\phi=e^{iK}$ and one recovers the differential form of the result of Ref.~\cite{0305132} that the integral of the K\"ahler form on a contractible cycle linking an A3-brane is equal to the linking number.

\subsection{Brane worldvolume actions}

The worldvolume action on the topological branes contains the term
\beq
S\supset\int_M \alpha\wedge j=\int_C \alpha
\eeq
where $C$ is the worldvolume of the brane, which is Poincar\'e dual to $j$.  When $j$ is the B2-brane current this reduces to the worldvolume action of Ref.~\cite{0012041}.  However it yields a much more general quantization condition, its exponential is only well-defined if the periods of the (real part of the) polyform $\phi$ on all (or half of all, if we choose a polarization) cycles are integral.

In addition topological branes carry $U(N)$ gauge bundles with curvature $F$ and so, ignoring the $\sqrt{\hat{A}}$ factor that is hidden in the delta function \cite{9710230}, one may conjecture, by analogy with D-brane Wess-Zumino terms, that the action is
\beq
S=\int_C \alpha\wedge\Tr{e^F}. \label{cs}
\eeq
In the case of the physical string theory, this is just a nonabelian generalization of the superpotential of Ref.~\cite{Luca}.  Such a guess is gauge-invariant under transformations of $B$ that preserve $B+F$.  One consequence of such an action would be that the A1-branes proposed in Ref.~\cite{0312022} may be realized as gauge instantons on coisotropic A5-branes.

As suggested in \cite{Luca,giulio} integrating by parts one obtains familiar Chern-Simons actions multiplied by components of $\phi$.  For example, for the B6-brane on a Calabi-Yau $\phi=\Om$ and so
\beq
S=\int_M \Om\wedge\Tr(A\wedge dA+\frac{2}{3}A\wedge A\wedge A)
\eeq
where the contraction with $\Om$ kills the $\partial$ part of $d$ leaving $\overline{\partial}$.  Notice that the quantization of $\Om$ in this case is equivalent to the usual condition that the level of the Chern-Simons theory be integral \cite{Nekrasov}.  Interestingly in the case of a probe coisotropic A5-brane on a type 0 generalized complex Calabi-Yau
\beq
\phi=\phi_0 e^{B+iK}
\eeq
and so the action contains the sum of the Chern-Simons 5-form, the wedge product of the Chern-Simons 3-form with the K\"ahler potential and the wedge product of the trace of the connection with the square of the K\"ahler form.  There are also couplings to the B-field in Eq.~(\ref{cs}) which naively lead to contributions to the lower brane charges as in \cite{0003037}.  In the physical string theory such charges are often canceled by bulk contributions \cite{0004141} to the action.  It seems plausible that the Hitchin functional (\ref{hitch}) contains terms that play a similar role in the topological theory.  

\section{Twisted homology and K-theory} \label{ksec}

\subsection{Differential forms}

If the B-field is not closed then one can define a 3-form $H=dB$.  The well-definedness of the fundamental string partition function $e^{iB}$ then implies that $H$ is quantized.  The possibility of adding $k$th roots of unity to the partition function of strings wrapped on $\Z_k$-valued torsion 2-homology cycles defines a lift of $H$ to integral cohomology.  For simplicity we will ignore these phases and consider $H$ to be a quantized differential form, commenting on the complications that arise in the integral cohomology case when they appear and in particular in Subsec.~\ref{torsec}.

The Hitchin functional is still given by Eq.~(\ref{hitch}) in the presence of a nontrivial $H$ flux, but the integrability condition changes.  We will restrict our attention to integrable generalized Calabi-Yau structures, which we have argued correspond to configurations without topological branes.  As in QED and physical string theories, while we are considering backgrounds without sources we will still impose that the fluxes are compatible with the existence of probe branes.  In other words, we will restrict our attention to quantized generalized Calabi-Yau structures, although it may be that for some physical applications such a restriction is unnecessary.

When the $H$ flux is nontrivial the integrability condition is twisted
\beq
(d+H\wedge)\phi=0 \label{int}
\eeq
and so locally one may define a potential $\alpha$ by
\beq
\phi=(d+H\wedge)\alpha
\eeq
which is defined up to a gauge transformation
\beq
\alpha\longrightarrow \alpha+(d+H\wedge)\alpha.
\eeq
The gauge invariance, and the fact that $\phi$ satisfies (\ref{int}), are consequences of the nilpotence of the differential $(d+H\wedge)$
\beq
(d+H\wedge)\phi=(d+H\wedge)(d+H\wedge)\alpha=d^2\alpha+d(H\wedge\alpha)+H\wedge d\alpha+H\wedge\H\wedge\alpha=0.
\eeq
Here the $d^2$ terms vanishes because $d$ is nilpotent, the $H\wedge\H\wedge\alpha$ terms vanishes because
\beq
H\wedge H=0 \label{hh}
\eeq
and the other two terms cancel using the fact that $H$ is closed and the Leibniz rule
\beq
d(H\wedge\alpha)=(dH)\wedge\alpha-H\wedge d\alpha.
\eeq
While (\ref{hh}) holds at the level of differential forms, in the full integral cohomology $H\cup H$ need not be zero, it may be any 2-torsion element of the sixth cohomology.  However in the case of a 6-dimensional orientable manifold the 6th cohomology group does not contain any torsion and so $H\cup H$ vanishes.

As is the case for Ramond-Ramond fields in the physical string theory, when $H\neq 0$ the gauge-invariant quantity $\phi$ is not quantized, in fact it would be impossible to quantize it since it's not closed.  Intuitively the quantity which is quantized is $e^{B}\phi$.  This quantity is closed, however as $B$ is not globally defined it requires further explanation.

To understand the twisted quantization condition more concretely, expand a type $k$ pure spinor $\phi$ in terms of forms of different degrees
\beq
\phi=\phi_k+\phi_{k+2}+\phi_{k+4}+\phi_{k+6}
\eeq
where $\phi_k$ is the lowest degree form component of the polyform $\phi$.  The integrability condition (\ref{int}) now implies
\beq
d\phi_k=0\hsp H\wedge\phi_j=d\phi_{j+2}\hsp H\wedge\phi_{k+6}=0 \label{ritorta}
\eeq
for all $j$.  The closure of the lowest component $\phi_k$ implies that $\phi_k$ may be the right term to quantize.  By analogy with the Page charges of the physical string \cite{0006117} we will assume that $\phi_k$ is quantized and see what the other conditions in (\ref{ritorta}) imply.  We will continue to treat $\phi_k$ as a quantized differential form, but bear in mind that it is really an element of integral cohomology.  In particular, we will identify backgrounds in which the value of $\phi_k$ differs by a exact form, as in this case one may chose the higher degree components such that $\phi$ differs by a $(d+H\wedge)$-exact form and so the Hitchin functional is unchanged. We will comment later on the effect of such a transformation on $\overline{\phi}$, which in principle may transform by a nonexact form and so change the Hitchin functional.  For now we will treat $\phi$ and $\overline{\phi}$ as independent variables, analogously to the fact that one does not impose RR self-duality when arriving at the K-theory classification of RR fields in the physical string theory.  One may object that the addition of an exact form to $\phi$ changes it and so we have no right to quotient by exact forms.  The addition of an exact form $d\alpha$ does change the worldvolume action of any topological branes, which contains a term $\int\alpha$, however we have already assumed that there are no branes in our configuration.

The next condition is
\beq
H\wedge\phi_k=d\phi_{k+2} \label{condk}
\eeq
which implies that $\phi_k$ cannot be an arbitrary element of cohomology, but rather it is in the kernel of the map $(H\wedge)$ which wedges an element of de Rham cohomology with $H$, since the right hand side is exact and so trivial in de Rham cohomology.

One particularly interesting case is $k=0$.  In this case $\phi_0$ is closed and so is constant and it is necessarily nonzero as $k=0$, Eq.~(\ref{condk}) implies
\beq
B=\Re(\phi_{2})/\Re(\phi_0).
\eeq
As $\phi_{k+2}$ is globally defined, so is $B$.  We have demonstrated that {\sl{in the absence of sources either the generalized Calabi-Yau structure of the A model is type 2 or else $H$ is exact.}}  In the presence of sources one instead finds that the A3-brane current is Poincar\'e dual to $\phi_0 H$.  Therefore the A model may exist on a space with a type 0 generalized Calabi-Yau structure and a nonexact $H$ flux, but the flux will be dual to an A3-brane worldvolume.  This is similar to Romans' supergravity, in which the product of the Romans mass and the $H$ flux is Poincar\'e dual to a D6-brane worldvolume.

If we add something in the image of $(H\wedge)$ to $\phi$
\beq
\phi\longrightarrow\phi+H\wedge\alpha
\eeq
then the $(d+H\wedge)$ closure of $\phi$ before and after the transformation implies that the new term again does not modify the action (\ref{hitch}).  Thus elements $\phi$ are classified by the kernel of $(H\wedge)$ quotiented by the image of $(H\wedge)$ in de Rham cohomology, in other words by the cohomology $\HHEO$ of the operator $(H\wedge)$
\beq
\phi\in\HHEO(M)=\frac{\Ker(H\wedge:\HEO\longrightarrow\HOE)}{\im(H\wedge:\HOE\longrightarrow\HEO)}.
\eeq

Now we are ready to impose the next condition in (\ref{ritorta})
\beq
H\wedge\phi_{k+2}=d\phi_{k+4}. \label{terza}
\eeq
This implies that even when (\ref{condk}) is satisfied for some $\phi_{k+2}$ because $\phi_k$ is in the kernel of $H\wedge$, the pure spinor $\phi$ may still fail to be integrable and so we need to impose another condition.  The condition that such a $\phi_{k+4}$ exist is the vanishing of an object called a Massey product as an element of $\HHEO$
\beq
[H,H,\phi_{k}]=0. \label{massey}
\eeq
The Massey product of three differential forms $a,\ b$\ and $c$ is defined by
\beq
[a,b,c]=a\wedge d^{-1}(b\wedge c)+d^{-1}(a\wedge b)\wedge c
\eeq
and so Eq.~(\ref{massey}) is easily checked
\beq
[H,H,\phi_{k}]=H\wedge d^{-1}(H\wedge\phi_k)=H\wedge d^{-1}d\phi_{k+2}=H\wedge\phi_{k+2}=d\phi_{k+4}
\eeq
where we have used the fact that $H\wedge H=0$ to eliminate the second term in the Massey product.  While in general $d^{-1}d$ of a form is only well-defined up to the addition of an arbitrary closed form $\alpha$, which would change the Massey product by $H\wedge\alpha$, in this case the Massey product is well-defined as an element of the cohomology $\HHEO$ because we have quotiented by all forms $H\wedge\alpha$.  Thus (\ref{massey}) is precisely the integrability condition (\ref{terza}).

As we are interested in 6-dimensional manifolds, the condition (\ref{massey}) is nearly always trivial.  The result is a $(k+5)$-form, and so it may only be nontrivial when $k=0$ or $k=1$.  If $k=0$, then we have already demonstrated that $H$ is exact.  Therefore one solution of (\ref{condk}) for $\phi_2$ is $\phi_2=\phi_0 B$.  One can find $\phi_4$ by imposing that its exterior derivative is the Massey product (\ref{massey}) which is equal to $H\wedge\phi_2=H\wedge B\wedge\phi_0$.  One solution is just half of $\phi_0B\wedge B$.  Thus for any $\phi_0$ one can find an integrable Calabi-Yau structure
\beq
\phi=\phi_0 e^B. \label{type0}
\eeq
These are not the only structures with a given $\phi_0$, as one can add $(d+H)$-closed forms with higher values of $k$.  Generally in doing so one ruins the purity of the spinor, which we have not yet imposed.  
In conclusion, it appears as though the Massey product (\ref{massey}) on 0-forms is always trivial, and so when $H$ is exact $\phi_0$ can always be extended to a pure spinor.

Now we turn to the action of (\ref{massey}) on 1-forms $\phi_1$ in the case $k=1$.  The Massey product
\beq
d\phi_5=[H,H,\phi_1]=H\wedge\phi_3
\eeq
must be a six-form, so it is proportional to the volume form.  The crucial observation is that the wedge product is nondegenerate on middle dimensional cohomology classes of orientable $(4k+2)$-manifolds, and in particular on 3-classes of 6-manifolds.  Therefore any 6-cocycle is the product of the cohomology class of $H$ with some other cohomology class.  At the level of differential forms this is the condition that any 6-form $A_6$ is equal to
\beq
A_6=H\wedge C_3+dB_5
\eeq
for some closed 3-form $C_3$ and some 5-form $B_5$.  $C_3$ is closed and so
\beq
\phi_3\p=\phi_3-C_3
\eeq
also satisfies (\ref{condk}).

We now choose
\beq
A_6=H\wedge\phi_3
\eeq
and use the corresponding $C_3$.  Then
\beq
H\wedge\phi_3\p=H\wedge\phi_3-H\wedge C=dB_5
\eeq
and so one may set
\beq
\phi_5=B_5.
\eeq
Therefore we have a constructed an integrable polyform
\beq
\Phi=\phi_1+(\phi_3-C_3)+B_5.
\eeq
In conclusion, the Massey product condition (\ref{massey}) is always satisfied for some choice of $\phi_3\p$ and $\phi_5$ given a $\phi_1$ and $\phi_3$ that satisfy (\ref{condk}).  Therefore both even and odd $(d+H)$-closed polyforms are arbitrary linear combinations of solutions of (\ref{condk}) with $\phi_0$ closed.


The next conditions in (\ref{ritorta}) increase the dimension of a form by more than 6 and so are trivial on a 6-manifold.  Thus we are at the end, and conclude that at the level of differential forms integral complex structures are classified by the cohomology of the operator $(H\wedge)$ in the de Rham cohomology.  In particular we have demonstrated

\noindent
{\it{Every quantized even/odd integrable generalized complex structure on an orientable 6-manifold $M$ uniquely defines an element of the free part of $\HHEO$.}}

The free part is the integral lattice of the corresponding de Rham cohomology, corresponding to differential forms in $\HHEO$ with quantized periods.  In general $\HHEO$ with integral coefficients will have a torsion part as well, which is critical to the identification with twisted K-theory and will be the subject of the next subsection.


\subsection{A twisted K-theory classification?} \label{torsec}
So far we have described the integrability condition for differential forms which may be taken to be quantized.  This corresponds to a classification of pure spinors via the integral lattice of de Rham cohomology.  In general however, once the quantization condition has been imposed, one is free to add phases to the path integrals of branes that wrap cycles representing torsion homology classes $\Z_k$.  Such phases have already appeared in the literature, an example being the exponentiated period of the 2-form $b$ on page 15 of \cite{0411073}.  The choice of these phases in addition to an element of the integral lattice of de Rham cohomology yields an element of cohomology with integral coefficients.  In the attractor picture it is clear that one needs to use integral cohomology and not simply an integral lattice in real cohomology, because the D-branes responsible for attractor flows may wrap any cycle in the dual integral homology, including those with torsion $(\Z_k)$ charges.  Once torsion cohomology classes are included, one needs to refine the multiplication $\wedge$ of differential forms to obtain the correct torsion part.  This refined product is denoted $\cup$.

At the level of integral cohomology things are a bit more complicated.  The condition (\ref{massey}) is replaced by an object called a Toda bracket, which does not necessarily vanish.  The Toda bracket is just the Massey product but instead of differential forms one uses simplicial chains, $d$ is the boundary operator and the wedge product is replaced by the intersection. Tensoring the cohomology classes by the real numbers the Toda bracket is equal to the Massey product.  Since we have shown that the Massey products vanish, the Toda bracket is in the kernel of this tensoring, which is the torsion part of the cohomology.  The Toda bracket
\beq
[H,H,x]=0 \label{toda}
\eeq
where $H$ corresponds to any codimension 3 chain which is Poincar\'e dual to $H$, augments the dimension of $x$ (seen as a cocycle) by 5, and so it gives either a 5-class if $x$ contains a 0-component or a 6-class if $x$ contains a 1-component.  A more pedagogical introduction to this operation in the context of physical string theory, with examples and pictures, can be found in Ref.~\cite{0611218}.

If there are worldsheet global anomalies of the kind introduced in Ref.~\cite{FW} then one may need to add a $\Z_2$ torsion correction called $Sq^3$ to the condition (\ref{toda}) as in Ref.~\cite{DMW}.  In addition one may also need an additional $\Z_3$-torsion operator which also augments the degree by 5, as in Ref.~\cite{representable}.  As both corrections yield torsion classes 5-dimensions higher, and since the top class contains no torsion, again they can only be nontrivial on the zero-class $\phi_0$.  Thus, such corrections only serve to augment the minimum quantum of $\phi_0$, and as a result the Dirac quantization of A5-branes.

In the B model $\phi$ is an odd polyform, so the lowest possible class is a 1-class and so the Toda bracket, even with the possible corrections noted above, is a torsion class in the sixth integral cohomology, which again is trivial for an orientable six-manifold.  Thus in the B model (\ref{toda}) is satisfied and quantized pure spinors are really classified by the $H\cup$ cohomology of the integral cohomology
\beq
\phi_{\rm B\ model}\in\HHO(M;\Z)=\frac{\Ker(H\cup:\HO\longrightarrow\HE)}{\im(H\cup:\HE\longrightarrow\HO)}.
\eeq
However some torsion values of the 5-class may be in the image of (\ref{massey}) and so do not contribute to an integral version of the Hitchin functional.

In the A model, on the other hand, if the background is type 0 then $\phi$ will contain a 0-form component.  The Toda bracket will be a torsion element of the fifth integral cohomology, which is isomorphic to the torsion part of the first homology.  Thus the Toda bracket will be zero if the first homology group contains no torsion, for example if the 6-manifold is simply connected.  The torsion part of the first homology group with integral coefficients is measured by the homology group with $U(1)$ coefficients.  
Therefore we have argued that

\noindent
{\it{If $\H_1(M;U(1))=0$ then a quantized generalized complex structure determines an element of $\HHEO(M;\Z)$ which is in this case, as a set, isomorphic to the twisted K-theory of $M$.}}


Ignoring the Toda bracket terms, in both the A and B model this sequence agrees, up to $Sq^3$ terms, with the Atiyah-Hirzebruch spectral sequence for twisted K-theory of Ref.~\cite{math/0510674}.  $Sq^3$ automatically vanishes on an orientable 6-manifold, and so we conclude that integrable, quantized pure spinors in both the A and B models on 6-manifolds are, as a set, classified by twisted K-theory as was conjectured by Hitchin in Ref.~\cite{math/0209099}. Thus when the Toda brackets are zero one may identify each background of the topological theory with a twisted K-theory class.  In the A model this identification holds even in the presence of Toda bracket terms, but one needs to scale the 0-form piece.  In the case of the B model the identification also continues to work when the Toda bracket term is nonzero, but needs to identify backgrounds which differ by those torsion characteristic 5-classes which are in the image of the Toda bracket.

While we have argued that for every allowed source-free background one can associate a twisted K-theory class, unlike the case of physical string theory the correspondence does not appear to work the other way.  One problem is that there are some twisted K-theory classes that do not correspond to any background.  In generalized complex geometry one imposes that the pure spinor is nondegenerate, in other words that the Hitchin functional is nonvanishing.  This is equivalent to assuming that the volume of the 6-manifold is nonzero.  This assumption implies for example that the degree of the lowest nonexact form component of $\phi$ is at most 3.  However there are twisted K-theory classes for which this is not satisfied, such as the zero class.

This leaves two possibilities, either one admits backgrounds with zero volume, or one accepts that there are twisted K-theory classes which do not correspond to allowed backgrounds.  It may well be that, even allowing for volumeless compactifications, some twisted K-theory classes correspond to spinors which are never pure and so are still not allowed.  This seems very likely.  For example, consider the A model on a type 0 manifold with $B=0$.  The purity of the spinor demands that $\phi$ be of the form $e^{iK}$.  While the argument for the quantization of $K\wedge K$ is more solid, in this section we have imagined that also the 2-form part $K$ is quantized.  This implies that such configurations are entirely classified by line bundles, so that the 4-class is just the square of the 2-class.  While this is consistent with the melting crystal picture, it is not the case for an arbitrary untwisted K-theory class.  In general the second Chern character is not simply the square of the first, there may be instantons.  And so it appears as though at least some twisted K-theory classes are really missing.

This is not surprising in light of the fact that the pure spinor condition
\beq
\phi\Gamma\phi=0
\eeq
is a quadratic and its solutions form a cone.  In particular, there is no additive structure on the space of pure spinors.  On the other hand twisted K-theory is an abelian group.  Thus it would be surprising if there were an isomorphism between pure spinors and twisted K-theory.  A similar problem arises in the physical string theory if one attempts to simultaneously classify RR and NSNS fluxes.  These are also related by quadratic Bianchi identities, and so one finds a nonlinear constraint on the fluxes whose solutions in general are unlikely to posses an additive structure \cite{DMW,sduality}.

Not only do there appear to be twisted K-theory classes that do not correspond to any background.  But also multiple backgrounds may correspond to the same twisted K-theory class.  For example one may add something exact to $\phi$ and so change the background but not the twisted K-theory class.  However in this case it is possible that no observables in the topological string theory are sensitive to the shift.  For example, the Hitchin functional is unchanged. However when $\phi$ changes by an exact form, its dual $\overline{\phi}$ may be shifted by a form which is not exact, which may change the Hitchin functional and so be detectable by the source-free topological theory.  Thus while it is clear that multiple backgrounds correspond to the same twisted K-theory class, it is not yet clear whether these multiple backgrounds correspond to identical physics.

Quantized generalized Calabi-Yau structures do not only appear in topological
string theory, but also may exist in physical string theories.  Ramond-Ramond
field strengths are quantized and in 6-dimensional compactifications
preserving $\mathcal{N}=1$ supersymmetry are pure.  Thus, when they
are nondegenerate, they yield quantized generalized Calabi-Yau structures.
Without imposing supersymmetry, Ramond-Ramond field strengths are classified
by twisted K-theory.  Supersymmetric configurations correspond to pure field
strengths and so may correspond to the same subset of twisted K-theory as the
quantized pure spinors in the topological theory.

\section{Type change from coisotropic branes} \label{jumpsec}
The integrability condition is particularly strong on the 0-form $\phi_0$.  Its closure implies that it is constant.  The quantization condition further implies that its real part is an integer.  In this section we will see that these two conditions imply that a particular phenomenon, known as type jumping, occurs in the A model only when crossing coisotropic branes.  This contrasts with the generalized complex case studied by Gualtieri in which it occurs on codimension 2 submanifolds and also with the unquantized generalized Calabi-Yau case in which it never occurs.

The type of a generalized complex manifold is the degree of the lowest component of its polyform pure spinor $\phi$.  As this polyform is even in the A model and odd in the B model, so is the type.  The nondegeneracy of the polyform, which is equivalent to the Hitchin functional or the volume of the 6-manifold not vanishing, implies that the degree is at most half of the dimension of the manifold.  Thus the A model may only compactified on a generalized complex manifold of type 0 or 2 and the B model on a manifold of type 1 or 3.

The A model and B model only have worldvolume formulations on manifolds of type 0 and 3 respectively, and so it is not known whether the A model can be defined on a type 2 background or the B model on a type 1.  It is even less clear whether both types may coexist in the same space.  The above considerations place some restrictions on the kind of type change that may be allowed in topological string theories, which will be the subject of this short section.

Gualtieri has demonstrated in Ref.~\cite{math/0401221} that the type of a generalized complex manifold is upper semi-continuous.  In other words, higher types generically occur on submanifolds of nontrivial codimension.  Generically a complex structure two types higher will occur on a submanifold of codimension two.  This is a simple consequence of the fact that a type $k$ even or odd complex polyform becomes type $k+2$ when the leading component vanishes, which is a single complex condition and therefore is satisfied on a codimension two submanifold.  Thus one might expect that in the A-model  a type 0 pure spinor might change to type 2 generically on a 4-dimensional submanifold, and similarly for type 1 changing into type 3 in the B model. How does this pattern change when the pure spinors are quantized?

In the B model one jumps from type 1 to type 3 when the 1-form part of $\phi$ vanishes, which is sourced by the B4-brane.  The B4-brane is of real codimension 2 and so is not a domain wall.  Instead, type 3 generically exists in type 1 on codimension 2 submanifolds as indicated by Gualtieri's analysis.  For example, the following nondegenerate, integrable generalized complex structure on the six-torus  satisfies the untwisted quantization condition
\beq
\phi=\textup{sin}(\Re(2\pi z_1))dz_1+dz_1\wedge dz_2\wedge dz_3 \label{type}
\eeq
where $z_i$ are complex coordinates for the torus such that the real parts are periodic modulo $1$.  This is type 1 almost everywhere but is type 3 on the subtorii at $z_1=0$ and $z_1=1/2$.  Note however that the 1-form component is exact, and so it does not contribute to the Hitchin functional.  Thus the theory, in the absence of branes, may be indistinguishable from a theory which is everywhere type 3.  However
\beq
\phi=2\textup{sin}^2(\Re(2\pi z_1))dz_1+dz_1\wedge dz_2\wedge dz_3
\eeq
exhibits the same type structure while the type 1 part is not exact.  Yet, this is cohomologous to a configuration which is everywhere type 1.  And so it is possible that type change is not detected by any topologically observable of the theory, at least in the absence of sources.

Let us now turn to the A model and consider the pure spinor
\beq
\phi = \textup{exp}((dz_1\wedge d{\bar z}_1 + dz_2\wedge d{\bar z}_2 + dz_3\wedge d{\bar z}_3)/2) \Bigl(m_1\,  \theta(\Re z_1) \theta(a- \Re z_1 )  + m_2 \, dz_1\wedge dz_2 \Bigr)
\eeq
where $m_1$ and $m_2$ are integers and  the step function $\theta(\Re z_1) = 1$ for  $\Re z_1 >  0$ and vanishes otherwise.  In the interval $0< \Re z_1 <  a$ the pure spinor is of  type 0 for SU(3)$\times$SU(3) structure on $TM\oplus T^*M$ and can be written as
\beq
\phi = m_1 e^{ \frac{1}{2} dz_3\wedge d{\bar z}_3 } \, \textup{exp}  \Bigl( ik
+ \frac{m_2}{m_1} dz_1\wedge dz_2 \Bigr) \label{awall}
\eeq
with $k= ( dz_2\wedge d{\bar z}_2 + dz_3\wedge d{\bar z}_3)/2$. The spinor is a standard type 2 elsewhere. The type change occurs as one crosses coisotropic (real codimension 1) branes with currents
\beq
d \phi = m_1 \, \Bigl( \delta^1(\Re z_1) - \delta^1 (a- \Re z_1 ) \Bigr)\textup{exp}((dz_1\wedge d{\bar z}_1 + dz_2\wedge d{\bar z}_2 + dz_3\wedge d{\bar z}_3)/2)
\eeq
yielding zero total brane charge.

A5-branes are obstructions to the integrability of the generalized
Calabi-Yau structure, and so we have seen that type change in the A model only
occurs where the pure spinor fails to be integrable.  Dirac quantization
implies that A5-branes are never smeared, they are localized on 5-dimensional
submanifolds and so type change occurs on sharp domain walls.  By
contrast, in the physical string theory the obstruction to integrability is
the RR field strength, which is almost always smeared throughout the
spacetime.  Thus in general in physical string theories one does not expect
domain walls separating regions of different types, instead regions of higher
type will generically occur on submanifolds in regions of lower type.  It is
possible to create a fat type-changing domain wall by hand by confining the
flux to a region with compact support.  In general such configurations will
not minimize energy unless there is some condensate which confines the flux.

One may try to construct type changing domain walls in the B model, which
separate domains at type 1 and type 3.  Such constructions can never be
consistent with the quantization condition, as a consequence of the fact that
there are no domain walls in the B model.  For example, one straightforward
generalization of the domain wall in Eq.~(\ref{awall}) would be the pure spinor
\beq
\phi = dz_3 \Bigl( \theta(\Re z_1) (\theta(a- \Re z_1 )  +   dz_1\wedge dz_2
\Bigr).
\eeq
The type of this pure spinor does jump on two domain walls, located at
$\Re(z_1)=a$ and $\Re(z_1)=0$.  However these walls are 5-dimensional and carry
A4-brane charge.  Thus they represent a smeared A4-brane solution and so are
inconsistent with pure spinor quantization.  For example, integrate $\phi$
over a loop which follows one wall in the $\Re(z_3)$ direction for a distance $L$,
then crosses the wall, then goes back on the other side of the wall, crosses
again and closes.  The integral will be $\pm L$, which may be any real number,
in contradiction with the quantization condition.  Thus the B0-brane partition
function is ill-defined in this background.

To summarize, Gualtieri's analysis for continuous generalized complex structures classified by complex forms continues to work for the B model (as well as the physical string theory, where $\phi$ is not quantized) and type change occurs on submanifolds of positive codimension and is  not sourced by branes.  The generalized Calabi-Yau and quantization conditions change this analysis for the A model.  As we saw in a specific example, the Bianchi identity (\ref{bianchi}) implies that the 0-form part of the A model pure spinor is an integer which can jump only when crossing a coisotropic A5-brane.  It is not a continuously varying complex number.
Thus in the A model type 0 and type 2 generalized complex structures exist on 6-dimensional open subsets of the spacetime separated by coisotropic branes. This is the only type change allowed in the A model.

\section{Conclusions}

In this note we have argued that the pure spinors which classify topological string theory backgrounds need to be quantized.  We have shown that in some cases such a quantization condition is necessary for the path integrals of topological branes to be well-defined, and thus is necessary for the nonperturbative definition of the theory.  We have discussed implications of this quantization condition, such as the fact that each background corresponds to a twisted K-theory class, although the inverse correspondence does not appear to hold.  Another implication is that type jumping in the A model does not occur as it would in the unquantized case, instead type 0 and type 2 generalized complex backgrounds are separated by coisotropic domain walls.

In Ref.~\cite{0312085} the authors found a number of relations between the B model on certain Calabi-Yau's and exactly solvable 2d CFTs and also matrix models.  To test the quantization conditions conjectured here, one could search for the corresponding condition in the 2-dimensional model.  In the simplest cases, when the matrix model partition function is just the exponential of the partition function, the quantization condition just seems to imply that the partition function is well-defined.  One might dream however that at least in some case one would discover a new constraint on the 2-dimensional theory.

In sufficiently supersymmetric compactifications it was demonstrated in Ref.~\cite{9807087} that attractive varieties admit complex multiplication.  Our claim that topological string theories necessarily live on attractive varieties then leads to several avenues for further research.  For example, one may attempt to generalize this complex multiplication to the generalized Calabi-Yau case.  The most ambitious goal would be, at least when the target space admits sufficient supersymmetry, to find a formulation of topological string theory in terms of this arithmatic structure.

These applications are a bit abstract, in part because definitions of topological strings on generalized complex manifolds are formal at best and so far rarely applied.  On the other hand, the quantization condition also holds in the usual Calabi-Yau case.  But does it have any meaningful consequences for calculations that interest people?

The A model and B model are used respectively to calculate prepotentials and superpotentials in supersymmetric gauge theories.  In an application to phenomenology, one matches the physical string parameters to the real world, giving some value of the pure spinor $\phi$ which is almost never integral.  Then one passes to the topological string to calculate the prepotential or superpotential.  If one believes the above arguments, the topological theory is almost never nonperturbatively well-defined for a given value of $\phi$.  This leaves three possibilities.

First, it may be that the nonperturbative well-definedness is irrelevant to these perturbative calculations, although amplitudes in configurations with topological branes are used in the calculations of superpotentials.  But it may be that these calculations work for some reason independent of the topological string theory, for example they may be derivable directly from anomalies in the field theory as in Ref.~\cite{0211170}.

Second, it may be that these calculations really have some finite error associated with the difference between $\phi$ and the nearest quantized form.  It seems unlikely that this would have gone undetected for so long.

Finally, it may be that the space of attractive generalized K\"ahler manifolds is dense in the space of generalized K\"ahler manifolds, as in the case of attractive 4-dimensional Calabi-Yau's in Ref.~\cite{9807087}. This would be an interesting result.

\section* {Acknowledgement}

We would like to acknowledge M. Aganagic, G. Bonelli, A. Brini, L. Martucci, A. Neitzke, T. Pantev, D. Persson and A. Tomasiello for insightful discussions.  R.M.
is supported in part by RTN contract  MRTN-CT-2004-005104 and
by ANR grant BLAN06-3-137168.


\end{document}